\documentstyle[graphicx,amsfonts,amssymb,11pt]{article}
\newcommand{\dfr}[2]{\frac {\displaystyle #1}{\displaystyle #2}}

\begin{document}
\title{Description of paramagnetic--spin glass transition in Edwards-Anderson model in terms of critical dynamics}
\author{M.~G.~Vasin\footnote{Corresponding author}}

\maketitle
\begin{abstract}
Possibility of description of the glass transition in terms of critical dynamics considering a hierarchy of the intermodal relaxation time is shown.
The generalized Vogel-Fulcher law for the system relaxation time is derived in terms of this approach. It is shown that the system satisfies the
fluctuating--dissipative theorem in case of the absence of the intermodal relaxation time hierarchy.
\end{abstract}

\section{ INTRODUCTION}
Despite of significant success in description of the dynamics of glass-transition, which they managed to attain by now, in particular in the
framework of the mode coupling theory, the idea of description of glass-transition within the scaling theory of phase transitions stays as very
attractive one. At first sight from the point of view of technical approaches such task does not present any complexity \cite{Vas}. However, attempts
of description of the phase transitions in disordered systems within the framework of the scaling approach faced serious difficulties connected with
the necessity of a deviation from the framework of usual representations of the fluctuation theory of phase transitions \cite{Doc}.

It is possible that part of these difficulties is caused by using of the static fluctuation theory of phase transitions and by the complexity of
averaging over configurations of the exchange integrals. In this work an attempt to consider this problem within dynamic approach, which is an
alternative to the replica method and in which such difficulties are absent, is made. It is assumed, what the ultrametric valleys space has a
hierarchical structure and is self-similar. Therefore it seems to be possible that we can formulate a dynamic scaling theory, in which, on the one
hand, the fundamentals would not be impaired, and, on the other hand, the system ultrametricity would be taken into account.

\section{THEORETICAL MODEL}
We will consider the Edwards-Anderson model close to the paramagnetic--spin glass transition temperature. Let us assume that the states space of the
frustrated system is decomposed into a set of the valleys divided between each other by energy barriers with a complex relief, and describe the
states of every valley with functions $\varphi_a (t,\,\bar r) $, where $a$ is the valley number. We will emanate from the standard formulation of the
stochastic dynamics problem, and let us suppose that the states, corresponding to various valleys, interact with each other. Then, according to the
mode coupling theory, the generalization of this problem to the case of the fields $\{\varphi_a (t,\,\bar r)\}$ system has the following form:
$$
\partial _t\varphi_a(t,\,\bar r)=-\sum\limits_b\alpha _{ab}\dfr{\delta F \{\varphi\} }{\delta \varphi_b(t,\,\bar r)}+\xi_a(t,\,\bar r), \qquad \langle \xi_a(t,\,\bar r)
\xi_b(t',\,\bar r')\rangle = 2\alpha_{ab}\delta (\bar r-\bar r')\delta(t-t').
$$
Here the indexes $a$ and $b$ mark the valleys (modes), $\alpha _{ab}$ are the inter-mode (inter-valley) coupling coefficients,
$$
F\{\varphi _a(t)\}=\int d^dr\left[ \dfr 12(\nabla \varphi _a(t,\,\bar r))^2+\dfr 12\left(\beta+\varrho J_a(\bar r)\right)\varphi _a^2(t,\,\bar
r)+\dfr 14\upsilon\varphi _a^4(t,\,\bar r)\right] ,
$$
where $J_a(\bar r)$ are random fields ($\langle J\rangle=0$). For the convenience let us turn from the operator $\alpha _{ab}$ to the inverse
operator $\bar \alpha _{ab}$, then the problem has the form:
$$
\sum\limits_b\bar\alpha _{ab}\partial _t\varphi_b(t,\,\bar r)=-\dfr{\delta F \{\varphi\} }{\delta \varphi_a(t,\,\bar r)}+\zeta_a(t,\,\bar r), \qquad
\langle \zeta_a(t,\,\bar r) \zeta_b(t',\,\bar r')\rangle = 2\bar \alpha_{ab}\delta (\bar r-\bar r')\delta(t-t').
$$
Let us make use of the method of reduction of the stochastic problem to the quantum-field model and write the full dynamic generating functional of
the function $\varphi_a(t,\,\bar r)$ in the form:
$$
\begin{array}{l} \Phi \{\varphi\}=\prod\limits_{t,\,\bar r,\,a}\delta \left(\sum\limits_b\bar\alpha
_{ab}\partial _t\varphi_b(t,\,\bar r)+\dfr{\delta F \{\varphi\} }{\delta \varphi_a(t,\,\bar
r)}-\zeta_a(t,\,\bar r)\right)\times\\[10pt]
\times\left|\det \left( \bar\alpha _{ab}\delta(\bar r-\bar r')\partial _t +\dfr{\delta ^2F\{\varphi\}}{\delta \varphi_a(t,\,\bar r)\delta
\varphi_b(t',\,\bar r')}\right)\right|.
\end{array}
$$
Using the Grassman variable algebra and the diagram technique it can show that in case of a purely dissipative problem the stochastic equation has
only one solution, therefore one can eliminate the absolute value of the determinant. This functional is the distribution function analog and <<cuts
out>> from the continual integral over fields $\varphi_a(t,\,\bar r)$ only the configurations which satisfies the original stochastic problem.
Integrating over the random field $\zeta $ one gets the statistical sum in the following form:
$$
\begin{array}{c}
\displaystyle Z=\left\langle {\int \left(\prod\limits_a D\varphi_aD\varphi '_a \right)
\exp{\left[S\{\varphi ,\,\varphi ',\,J\}\right]}}\right\rangle_J,\\[10pt]
\displaystyle S\{\varphi, \,\varphi ',\,J\}=\sum\limits_{ab}\int dt\,d^dr\left[
\bar\alpha_{ab}\varphi '_a(t,\,\bar r)\varphi '_b(t,\,\bar r)-\bar\alpha_{ab}\varphi '_a(t,\,\bar
r)\partial_t\varphi _b(t,\,\bar r)+\right.\\[10pt]
+\left.\varphi '_a(t,\,\bar r)\nabla ^2\varphi _a(t,\,\bar r)-\beta\varphi '_a(t,\,\bar r)\varphi _a(t,\,\bar r)-\varrho J_a(\bar r)\varphi
'_a(t,\,\bar r)\varphi _a(t,\,\bar r)-\upsilon\varphi '_a(t,\,\bar r)\varphi _a^3(t,\,\bar r)\right] .
\end{array}
$$

Ultrametricity of the valleys space (and the time scale accordingly) is an important stage of our model definition. We do it by means of the random
<<impurity>> field $J$: Let us assume that the valley states are independent of one another, therefore the averaging over fields $J$ is carried out
for every valley independently --- the correlation of $J_a$ and $J_b$ is absent. However, upon expiration of $\tau_{ab}$ time, which is necessary for
the valleys $a$ and $b$ to come to equilibrium, their states cease to be independent and the averaging over $J$ becomes to be common. Thus, let us
assume that a correlator $\langle J_a(t,\,\bar r)J_b(t',\,\bar r')\rangle =0$, if the time gap between the replica states is smaller than some
characteristic time $\tau_{ab}$, which is necessary for the valleys $a$ and $b$ to come to equilibrium ($|t-t'|<\tau_{ab}$), and, otherwise, $\langle
J_a(t,\,\bar r)J_b(t',\,\bar r')\rangle =\delta (\bar r-\bar r')$ for $|t-t'|>\tau_{ab}$ (Fig.1). For that we represent the correlator of the random
bonds as follows:
\begin{figure}[h]
   \centering
   \includegraphics[scale=1]{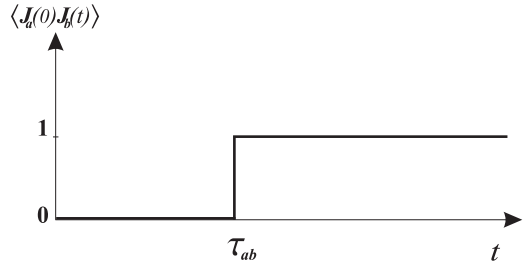}
   \caption{Graphic expression of $\langle
    J_a(t)J_b(t')\rangle $ correlator.}
\end{figure}
$$
\langle J_a(t,\,\bar r)J_b(t',\,\bar r')\rangle =\theta (|t-t'|-\tau_{ab})\delta(\bar r-\bar r').
$$
The relaxation time $\tau _{ab}$ depends on the distance apart the valleys $a$ and $b$ in the ultrametric space $x_{ab}$: $\tau_{ab}=\tau_0
e^{x_{ab}}$.

\section{RENORMALIZATION AND ANALYSIS}
Correlation functions of the fields, available in the model, have the following form, if $\bar\alpha_{aa}=\alpha_{aa}=1$,:
$$
\langle\varphi_a'\varphi_b\rangle =\dfr{\delta_{ab}}{i\omega +\varepsilon_k}, \qquad \langle\varphi_a\varphi_b\rangle
=\dfr{2\bar\alpha_{ab}}{\omega^2+\varepsilon_k^2}, \qquad \langle J_aJ_b\rangle =\dfr{e^{-i\omega \tau_{ab}}}{i\omega },
$$
where $\varepsilon_k=k^2+\beta$, and it was assumed that the <<inoculating>> operator,
$\bar{\alpha}_{ab}$, had the form $\bar{\alpha}_{ab}\simeq\delta_{ab}$. In this case the
contribution of $\varphi '_a(t,\,\bar r)\partial_t\varphi _b(t,\,\bar r)$ ($a\neq b$) is
unrenormalizable. Let us represent these correlation functions in form of the graphs a, b, and c
accordingly (Fig.2).
\begin{figure}[h]
   \centering
   \includegraphics[scale=0.7]{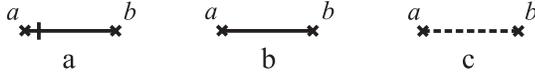}
   \caption{Graph representation of the correlators.}
\end{figure}
It can be checked that the formulated model is multiplicatively renormalizable, and the renormalized effective action has the following form:
$$
\begin{array}{c}
\displaystyle S^{(R)}\{\varphi, \,\varphi ',\,J\}=\sum\limits_{ab}\int d\omega\,d^dk\left[ Z_1\varphi '_a\varphi '_b+Z_2\omega\varphi '_a\varphi _b+\right.\\
\displaystyle-\left.Z_3k^2\varphi '_a\varphi _a-Z_4\varphi '_a\varphi _a-Z_5J_a\varphi '_a\varphi _a-Z_6\varphi '_a\varphi _a^3-Z_7i\omega e^{i\omega
\tau_{ab}}J_aJ_b\right] ,
\end{array}
$$
where $Z_1=Z_{\bar\alpha_{ab}}Z_{\varphi '}^2$, $Z_2=Z_{\bar\alpha_{ab}}Z_{\varphi '}Z_{\varphi }$, $Z_3Z_{\varphi '}Z_{\varphi }$,
$Z_4=Z_{\beta}Z_{\varphi '}Z_{\varphi }$, $Z_5=Z_{\varrho}Z_{J}Z_{\varphi '}Z_{\varphi }$, $Z_6=Z_{\upsilon}Z_{\varphi '}Z_{\varphi }^3$,
$Z_7=Z_{\chi}Z_{J}^2$ are corresponding renormalization constants. The canonical dimensions of the fields and parameters of this model are given in
the table:\\
\begin{center}
\begin{tabular}{|c|c|c|c|c|c|c|c|c|c|}
  \hline
   $F$ & $k$, $\nabla $ & $\omega$, $\partial_t$ & $\bar\alpha_{ab}$ & $\varphi$ & $\varphi '$ & $J$ & $\beta$ & $\varrho$ & $\upsilon$\\
  \hline
  $d_F^k$ & 1 & 0 & 2 & $-(1+d/2)$ & $-(1+d/2)$ & $-d/2$ & 2 & $2-d/2$ & $4-d$\\
  \hline
  $d_F^{\omega}$ & 0 & 1 & -1 & -1 & 0 & -1 & 0 & 0 & 0 \\
  \hline
  $d_F$ & 1 & z=2 & 0 & $-(3+d/2)$ & $-(1+d/2)$ & $-(2+d/2)$ & 2 & $2-d/2$ & $4-d$\\
  \hline
\end{tabular}
\end{center}

\begin{figure}[h]
   \centering
   \includegraphics[scale=0.7]{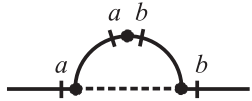}
   \caption{Graph which contributes to the renormalization of the term $\bar\alpha _{ab}$ (one-loop approximation).}
\end{figure}

The renormalization procedure is of fundamental importance.  The renormalization of the parameter $\bar\alpha _{ab}$ is a specially important one,
which looks as follows (Fig.3):
\begin{equation}\label{1}
\displaystyle Z_{\bar\alpha_{ab}}=\displaystyle \bar\alpha_{ab}-\varrho^2\dfr{2}{2!}\int\limits_{\lambda^z\omega_0<|\omega |<\omega_0}d\omega
\int\limits_{\lambda k_0<|k|< k_0}\dfr{d^dk}{(2\pi)^d}\dfr{\alpha_{ab}\omega^2+\bar\alpha_{ab}\varepsilon_k^2}{(\omega^2+\varepsilon_k^2)^2} \cdot
\dfr{e^{-i\omega \tau_{ab}}}{i\omega }.
\end{equation}
Here and below we amount to nothing more than the one-loop approximation. The integral over $\omega $ is divided into two parts: the first part is a
circulation integral around the pole $\omega =\varepsilon _k$, the second part --- around the pole $\omega =0$. As usual \cite{Halp}, in the first
case the integration is carried out over contour bounding of the complex space upper half-plane, whereas in the second integral, as well as when
integrating over $k$ we have to introduce the regularizing parameter $\lambda ^z$ ($z=2$ is the dynamics exponent). Thus, after the integration over
$\omega$ the expression (\ref{1}) has the form:
$$
\begin{array}{rl}
\displaystyle Z_{\bar\alpha_{ab}}\simeq\displaystyle \bar\alpha_{ab}-2\pi\int\limits_{\lambda k_0<|k|< k_0}\dfr{d^dk}{(2\pi
)^{d}}&\displaystyle\left[ \dfr {\varrho^2}2(\alpha_{ab}-\bar\alpha_{ab})\dfr{e^{\varepsilon_k\tau_{ab}}}{\varepsilon_k^2}-\dfr
{\varrho^2}4(\alpha_{ab}-\bar\alpha_{ab})\tau_{ab}\dfr{e^{\varepsilon_k\tau_{ab}}}{\varepsilon_k}\right.\\[10pt]
&\displaystyle\left.-\dfr {\varrho^2}2\dfr{\alpha_{ab}e^{\varepsilon_k\tau_{ab}}}{\varepsilon_k^2}-\dfr
{\varrho^2(1-\lambda^z)}{\pi}\dfr{\tilde{\tau}_{ab}\bar\alpha_{ab}}{\varepsilon_k^2}\right].
\end{array}
$$
Here the last term is the result of expansion in series of the exponential curve, besides the nondimensional time parameter $\tilde{\tau}_{ab}=\gamma
e^{\sigma x_{ab}}$ ($\gamma =\omega_0\tau_0$) is introduced. If one keeps in this formula only substantial logarithmically divergent terms, it has
the following form:
\begin{equation}\label{2}
\displaystyle Z_{\bar\alpha_{ab}}=\displaystyle\bar\alpha_{ab}+\bar\alpha_{ab}\varrho^2\left(\dfr{\tilde{\tau}_{ab}{(1-\lambda^z)}}{\pi}
+\dfr{1}{2}\right)\dfr{1}{2\pi }\ln \left(\dfr 1{\lambda }\right).
\end{equation}
Time ultrametricity at the integration over $\omega $ leads to the nontrivial dependance of the $\bar\alpha_{ab}$ renormalization on the $\lambda $.
Therefore, this very parameter has the most apparent features that are caused by the presence of the hierarchy of system relaxation times and have an
effect on the system dynamics.

$\beta$-renormalisation looks in the following form (Fig.4):
$$
Z_{\beta}\delta_{ab}=\displaystyle \beta\delta_{ab}-{\Sigma_1}_{ab}-{\Sigma_2}_{ab},
$$
where ${\Sigma_1}_{ab}$ ¨ ${\Sigma_2}_{ab}$ are contributions to $\beta $-renormalisation of (a) and (b) graphs, accordingly. The contribution of the
first graph has the form:
\begin{figure}[h]
   \centering
   \includegraphics[scale=0.7]{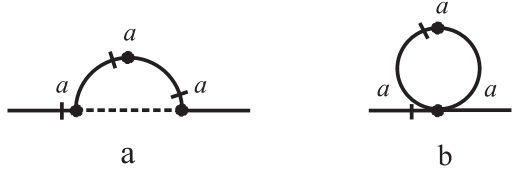}
   \caption{Graphs which contribute to the renormalization of the term $\beta $ (one-loop approximation).}
\end{figure}
$$
{\Sigma_1}_{ab}=\beta\varrho^2\dfr{2}{2!}\sum\limits_{c}\int\limits_{\lambda^z\omega_0<|\omega |<\omega_0}d\omega\int\limits_{\lambda k_0<|k|< k_0}
\dfr{d^dk}{(2\pi)^d}\dfr{-\alpha_{ac}\alpha_{cb}\omega^2+i2\alpha _{ac}\delta_{cb}\omega \varepsilon
_k+\delta_{ac}\delta_{cb}\varepsilon^2_k}{(\omega^2+\varepsilon_k^2)^2}\cdot \dfr{e^{-i\omega \tau_{ab}}}{i\omega }.
$$
Since $\bar\alpha_{ab}$ ($a\neq b$) parameter is divergent, the parameter $\alpha_{ab}$ will be small, respectively; therefore, only second and third
terms remain in the integral. After the integration over $\omega $, assuming that $\tau_{aa}=0$, we get the following:
$$
{\Sigma_1}_{ab}=-\delta_{ab}\dfr {\beta\varrho^2}2(1-\alpha_{ab})\int\limits_{\lambda k_0<|k|< k_0} \dfr{d^dk}{(2\pi)^{d-1}}\dfr
1{\varepsilon_k^2}=-\beta\varrho^2\delta_{ab}(1-\alpha_{ab})\dfr{1}{4\pi}\ln \left(\dfr 1{\lambda }\right)=0.
$$
I.\,e. in case of $\alpha_{aa}=1$, this graph does not contribute to renormalisation. The second cunterterm has the form:
$$
 {\Sigma_2}_{ab}= \dfr{\beta\upsilon\delta_{ab}(9\bar\alpha_{ab}+3\alpha_{ab})}{8\pi}\ln \left(\dfr 1{\lambda }\right).
$$
Thus, when $\alpha_{aa}=\bar\alpha_{aa}=1$, we have
$$
Z_{\beta}=\displaystyle \beta-\beta\dfr{3\upsilon }{2\pi}\ln \left(\dfr 1{\lambda }\right).
$$

The graphs renormalizing the point $\varrho\varphi_a ' \varphi_a J_a$ have the form represented in Fig.5.
\begin{figure}[h]
   \centering
   \includegraphics[scale=0.7]{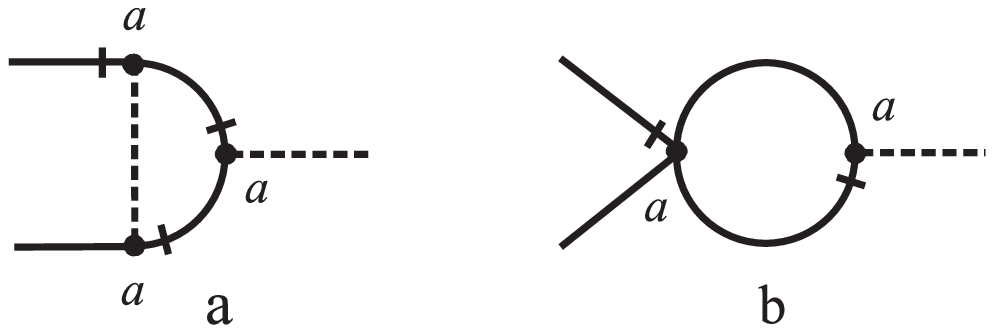}
   \caption{Graphs which contribute to the renormalization of the term $\varrho $ (one-loop approximation).}
\end{figure}
Using the above results it is simply to show that renormalisation of this point has the following form:
$$
Z_{\varrho}=\varrho-\varrho\dfr{3\upsilon }{2\pi}\ln \left(\dfr 1{\lambda }\right).
$$

In one-loop approximation the graphs renormalizing the point $\upsilon\varphi_a ' \varphi_a^3$ have the form represented in Fig.6.
\begin{figure}[h]
   \centering
   \includegraphics[scale=0.7]{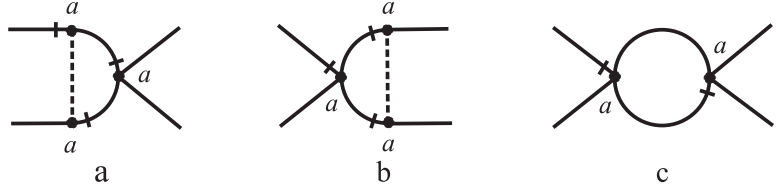}
   \caption{Graphs which contribute to the renormalization of the term $\upsilon $.}
\end{figure}
However, as it was shown above, the first graph (a) makes a zero-order contribution to renormalisation ($\Upsilon_1=0$). The second graph (b) makes
the following contribution:
$$
\Upsilon_2=\upsilon\varrho^2\dfr{6}{2!}\sum\limits_{c}\int\limits_{\lambda^z\omega_0<|\omega |<\omega_0}d\omega\int\limits_{\lambda k_0<|k|< k_0}
\dfr{d^dk}{(2\pi)^d}\dfr{\alpha_{ac}\alpha_{cb}\omega^2+\delta_{ac}\delta_{cb}\varepsilon^2_k} {(\omega^2+\varepsilon_k^2)^2}\cdot \dfr{e^{-i\omega
\tau_{ab}}}{i\omega },
$$
which has the following form after being integrated:
$$
\Upsilon_2=-\upsilon\varrho^23\pi\int\limits_{\lambda k_0<|k|< k_0} \dfr{d^dk}{(2\pi)^d}\dfr 1{\varepsilon_k^2}=-\upsilon\varrho^2\dfr{3}{4\pi}\ln
\left(\dfr 1{\lambda }\right).
$$
The third graph (c) contribution is
$$
\begin{array}{rcl}
 \Upsilon_3&=&\displaystyle 36\upsilon^2\delta_{ab}\sum\limits_{c}\int\limits_{\lambda^z\omega_0<|\omega
|<\omega_0}d\omega\int\limits_{\lambda k_0<|k|< k_0} \dfr{d^dk}{(2\pi)^d}\dfr{(-i\alpha_{ac}\omega
+\delta_{ac}\varepsilon_k)(\alpha_{cb}\omega^2+\bar\alpha_{cb}\varepsilon_k^2)}
{(\omega^2+\varepsilon_k^2)^3}=\\[10pt]
&=&\displaystyle\upsilon^2\dfr{9}{2\pi}\ln \left(\dfr 1{\lambda }\right),
\end{array}
$$
thus, we get:
$$
Z_{\upsilon}=Z_{\upsilon}=\upsilon-\Upsilon_1-\Upsilon_2-\Upsilon_3=\upsilon+\upsilon\varrho^2\dfr{3}{8\pi^2}\ln \left(\dfr 1{\lambda
}\right)-\upsilon^2\dfr{9}{2\pi}\ln \left(\dfr 1{\lambda }\right).
$$

Assuming that $Z_{\varphi}=\theta(\lambda)$, $Z_{\varphi '}=\theta '(\lambda)$, $Z_{J}=\vartheta (\lambda)$, and carrying out the scaling
transformation we get the following renormalization group:
$$
\begin{array}{rcl}
\bar\alpha_{ab}^{(R)}&=&Z_1\lambda^{d+z}=\displaystyle{\theta '}^2(\lambda
)\lambda^{d+z}\left[\bar\alpha_{ab}+\bar\alpha_{ab}\varrho^2\left(\dfr{\tilde{\tau}_{ab}(1-\lambda^z)}{\pi} +\dfr{1}{2}\right)\dfr{1}{2\pi}\ln
\left(\dfr 1{\lambda }\right)\right],
\\[10pt]
\bar\alpha_{ab}^{(R)}&=&Z_2\lambda^{d+2z}=\displaystyle\theta '(\lambda )\theta(\lambda )\lambda^{d+2z}\left[\bar\alpha_{ab}-O(\upsilon ^2)\right],
\\[10pt]
1&=&Z_3\lambda^{d+z+2}=\displaystyle\theta '(\lambda )\theta(\lambda )\lambda^{d+z+2},
\\[10pt]
\beta^{(R)}&=&Z_4\lambda^{d+z}=\displaystyle\theta '(\lambda )\theta(\lambda )\lambda^{d+z}\left[ \beta-\beta\dfr{3\upsilon }{2\pi}\ln \left(\dfr
1{\lambda }\right)\right],
\\[10pt]
\varrho^{(R)}&=&Z_5\lambda^{2d+2z}=\displaystyle\displaystyle\theta '(\lambda )\theta(\lambda )\vartheta(\lambda
)\lambda^{2d+2z}\left[\varrho-\varrho\dfr{3\upsilon }{2\pi}\ln \left(\dfr 1{\lambda }\right)\right],
\\[10pt]
\upsilon^{(R)}&=&Z_6\lambda^{3d+3z}=\displaystyle\displaystyle\theta '(\lambda )\theta ^3(\lambda
)\lambda^{3d+3z}\left[\upsilon+\upsilon\varrho^2\dfr{3}{4\pi}\ln \left(\dfr 1{\lambda
}\right)-\upsilon^2\dfr{9}{2\pi}\ln \left(\dfr 1{\lambda }\right)\right],\\[10pt]
\chi^{(R)}&=&Z_7\lambda ^{d+2z}=\displaystyle \vartheta ^2(\lambda)\lambda ^{d+2z}\chi.
\end{array}
$$
Taking into account canonical dimensions of the fields and introducing new denotations: $\xi =\ln (1/\lambda )$, $\varepsilon =4-d$, let us rewrite
this system in the following form:
$$
\begin{array}{rcl}
\bar\alpha_{ab}^{(R)}&=&\displaystyle\bar\alpha_{ab}+\bar\alpha_{ab}\varrho^2\left(\dfr{\tilde{\tau}_{ab}(1-e^{-z\xi })}{\pi}
+\dfr{1}{2}\right)\dfr{1}{2\pi}\xi ,
\\[10pt]
\beta^{(R)}&=&\displaystyle e^{2\xi}\left[ \beta-\beta\dfr{3\upsilon }{2\pi}\xi\right],
\\[10pt]
\varrho^{(R)}&=&\displaystyle e^{\varepsilon\xi/2}\left[\varrho-\varrho\dfr{3\upsilon }{2\pi}\xi\right],
\\[10pt]
\upsilon^{(R)}&=&\displaystyle e ^{\varepsilon\xi}\left[\upsilon+\upsilon\varrho^2\dfr{3}{4\pi}\xi-\upsilon^2\dfr{9}{2\pi}\xi\right].
\end{array}
$$
Let us expand exponents in series and keep only linear components in respect to $\xi $-terms. Supposing the continuity of the renormalization group
transformation let us describe the evolution of effective action parameters in the form of differential equations:
$$
\begin{array}{rl}
&\displaystyle\dfr{d\ln(\bar\alpha_{ab})}{d\xi}=\displaystyle\dfr{\varrho^2}{2\pi}\left(\dfr{\tilde{\tau}_{ab}(1-e^{-z\xi })}{\pi}
+\dfr{1}{2}\right),
\\[10pt]
&\displaystyle\dfr{d\ln(\beta)}{d\xi}=\displaystyle 2-\dfr{3\upsilon }{2\pi},
\\[10pt]
&\displaystyle\dfr{d\ln(\varrho)}{d\xi}=\displaystyle \dfr{\varepsilon}2-\dfr{3\upsilon }{2\pi},
\\[10pt]
&\displaystyle\dfr{d\ln(\upsilon)}{d\xi}=\displaystyle \varepsilon+\varrho^2\dfr{3}{4\pi}-\upsilon\dfr{9}{2\pi}.
\end{array}
$$
The fixed point condition is assigned by a set of equations: $\displaystyle\dfr{d\ln(\varrho)}{d\xi}=0$, $\displaystyle\dfr{d\ln(\upsilon)}{d\xi}=0$,
from which we get: $\displaystyle\upsilon^*=\dfr{\pi\varepsilon}{3}$, $\displaystyle\varrho^*=\sqrt{\dfr {2\pi\varepsilon}3}$. Then
$$
\begin{array}{rcl}
\beta^{1/2}&\simeq&\displaystyle\left(\dfr{T-T_c}{\kappa T_c}\right)^{1/2}=\displaystyle e^{\xi},\\[10pt]
\bar\alpha_{ab}&\simeq&\displaystyle\exp \left( \dfr{\varrho^2}{2\pi}\left(\dfr {\tilde{\tau}_{ab}}{\pi}+\dfr 12\right)\xi\right)\cdot \exp
\left(\dfr
{\varrho^2\tilde{\tau}_{ab}}{2\pi ^2z}e^{-z\xi}\right)=\\[10pt]
&=&\displaystyle\left(\dfr{T-T_c}{\kappa T_c}\right)^{\displaystyle\frac{\varepsilon}{3}\left(\dfr {\tilde{\tau}_{ab}}{\pi}+\dfr 12\right)}\cdot \exp
\left({\dfr {\varepsilon\tilde{\tau}_{ab}}{6\pi }\left(\dfr {\kappa T_c}{T-T_c}\right)^{z/2}}\right).
\end{array}
$$
Within continuous limit ($\bar\alpha_{ab}\to \bar\alpha (x)$) we can represent a temporary system evolution as the following functional relation:
$$
\displaystyle S(t)=\int dx \Delta '(x)\exp \left(-\dfr{t}{\bar\alpha (x)}\right),
$$
where $\Delta '(x)\sim j^{-x}$ is the distribution density of the valley pairs with respect to
distances, $x$, between such valleys in the ultrametric space. Hence, we can get the following
formula for the observable time of relaxation of a three-dimensional system:
$$
\displaystyle t_{rel}\sim\left(\dfr{T-T_c}{\kappa T_c}\right)^{\displaystyle\frac 13\left(\dfr {q}{\pi}+\dfr 12\right)}\cdot \exp \left({\dfr
{q\kappa}{6\pi}\dfr { T_c}{(T-T_c)}}\right),
$$
where $q=\gamma e^{\sigma/\ln j}$ is a value which is determined by ultrametric space parameters. Thus, on condition that $(T-T_c)\ll 1$, we come to
the Vogel-Fulcher-Tammann  law.

We would like to note that in a number of works the idea that critical phenomena in disordered systems can be described by assumption of exsistence
of the infinite continuous hierarchy of the correlation length and critical indexes was proposed (see, for example, \cite{Doc, Lud}). In our case the
renormalization (2) can be interpreted as the presence of such hierarchy.

Finally, let us write down the dynamic sensitivity of the system:
$$
G(\omega)=\sum\limits_{ab}\dfr{\bar\alpha_{ab}^{-1}}{i\omega+\bar\alpha_{ab}^{-1}\varepsilon _k}.
$$
In continuous limit this formula can be written in terms of <<intervalley transitions>> $x$:
$$
G(\omega)=\int\limits_{0}^{\infty }dx \Delta '(x)\dfr{\Gamma (x)}{i\omega+\Gamma (x)\varepsilon _k},
$$
where $\Gamma (x_{ab})=\bar\alpha_{ab}^{-1}$ is a value which is inverse to transition time between
the $a$ and $b$ valleys. Within $\omega \to 0$ limit it can be written down in a more traditional
form \cite{Gins}:
$$
D(\omega)\simeq \int\limits_{0}^{\infty }dx q'(x)\dfr{\Gamma (x)}{\omega ^2+\Gamma ^2(x)\varepsilon _k^2},
$$
where $q'(x)=\Delta '(x)/\Gamma (x)$. As is known  \cite{Gins}, to work the
fluctuating--dissipative theorem it is necessary that $q'(x)=\Delta '(z)$. Thus, this theorem is
valid when $\Gamma (x)=1$ at all $x$-values, i.e. in case of any whatsoever transition times
hierarchy is absent in the system.

\section{CONCLUSIONS}
The outcomes of this work has pointed out to a critical opportunity of application of the standard
methods of critical dynamics in description of glass transition. However, the given approach has
some weak-points. First of all, it relates to artificial introduction of the hierarchy of
relaxation times, therefore two scales of relaxation time do exist in this model: $ \bar\alpha $
corresponding to relaxation in the $\varphi $-fields subsystem; and $ \tau $ corresponding to the
$J $-fields subsystem, where the second one is entered to the model artificially. In fact these
time scales should be interconnected with each other, since relaxation of the $ \varphi $-fields
determines relaxation of the $J$-fields. Apparently, a form this interconnection directly depends
on a specific concrete physical problem, but in our model this is not taken into account. As a
consequence, $q$ does not depend on the temperature, that, apparently, is not exactly correct.
Therefore, it is possible to believe, that the obtained Vogel-Fulcher-Tammann formula is true in
case of weak temperature dependence of $q$.

\section{ACKNOWLEDGMENTS}
This study was supported by the RFBR grant (04-03-96020-r2004ural).


\end{document}